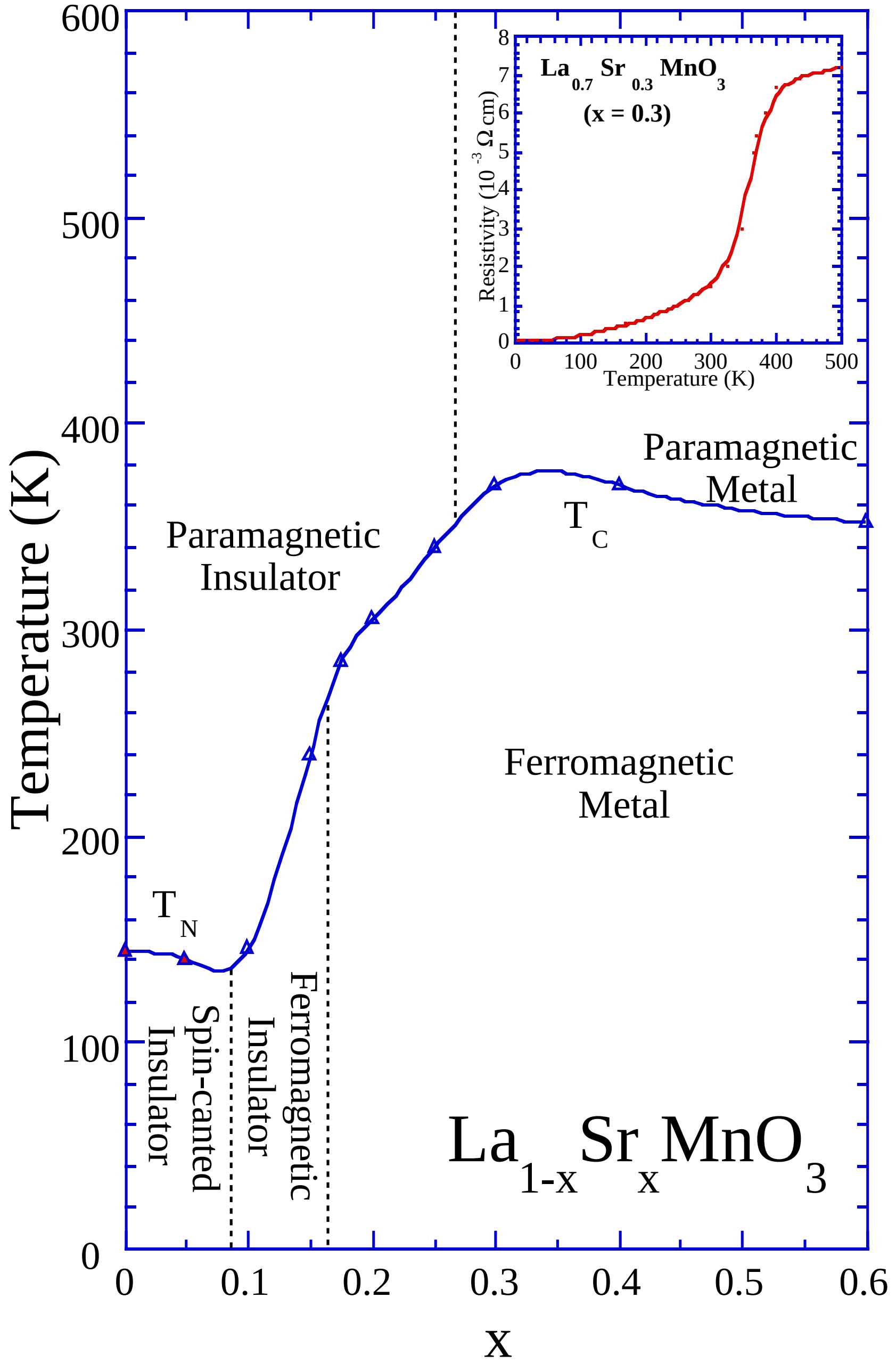

M.C. Martin *et al.*, Figure 1

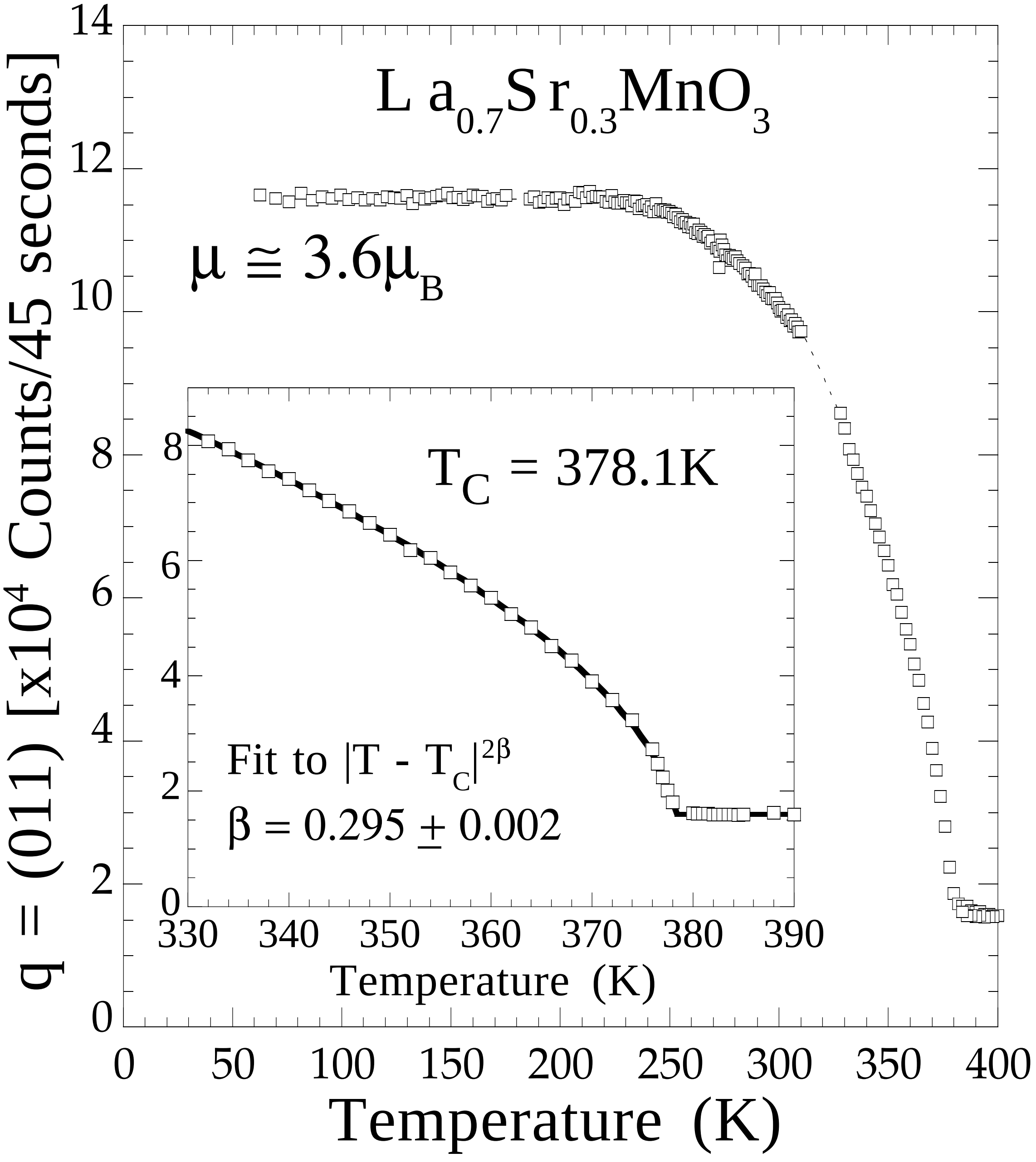

M.C. Martin et al., Figure 2

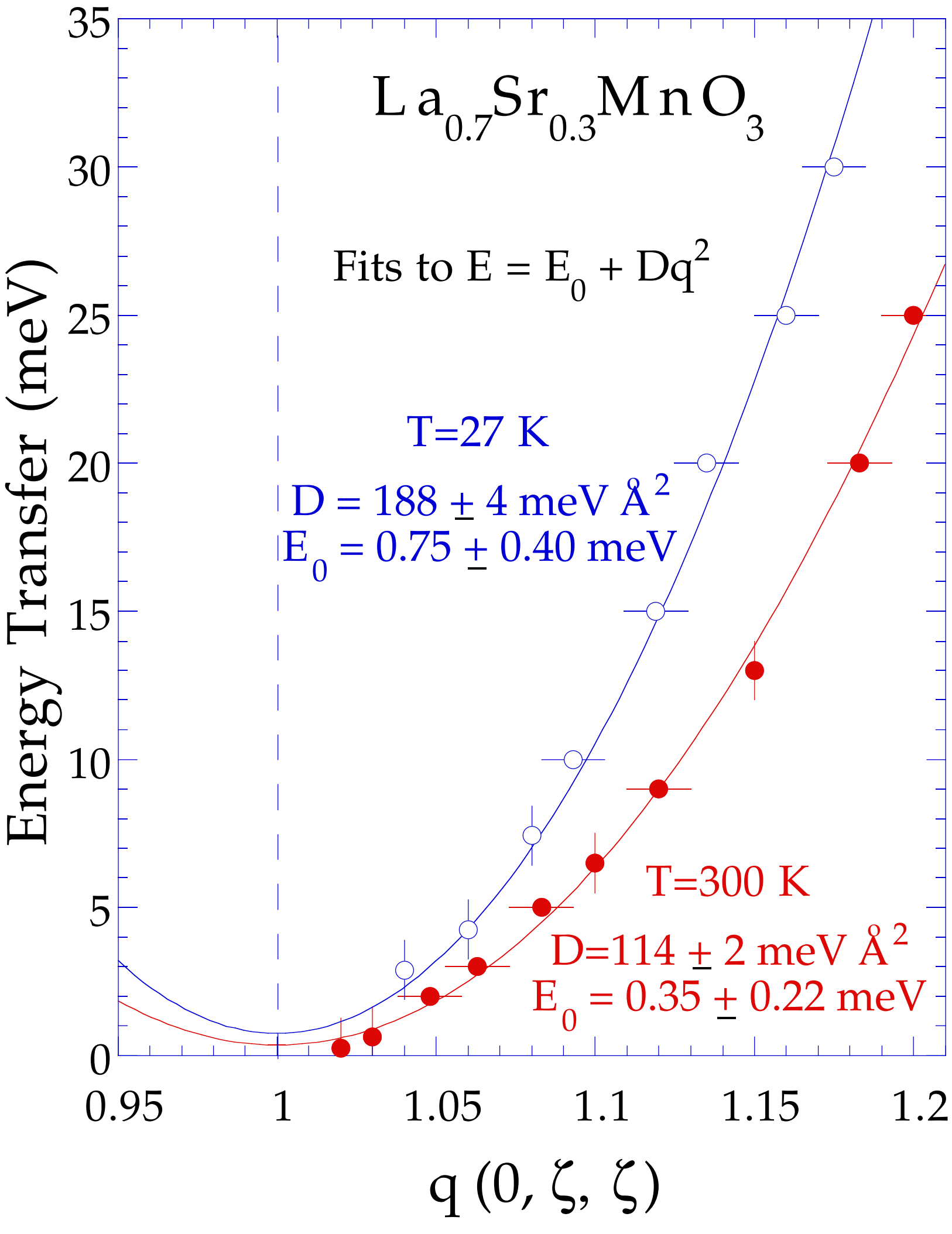

M.C. Martin et al., Figure 3

# MAGNETISM AND STRUCTURAL DISTORTION IN THE $La_{0.7}Sr_{0.3}MnO_3$ METALLIC FERROMAGNET

Michael C. Martin and G. Shirane
Department of Physics, Brookhaven National Laboratory, Upton NY 11973

Y. Endoh and K. Hirota
Department of Physics, Tohoku University, Sendai 980, Japan

Y. Moritomo and Y. Tokura
Joint Research Center for Atom Technology, Tsukuba, Ibaraki 305, Japan
and
Department of Applied Physics, University of Tokyo, Tokyo 113, Japan

## ABSTRACT

Neutron scattering studies on a single crystal of the highly-correlated electron system, $La_{1-x}Sr_xMnO_3$ with x ≈ 0.3, have been carried out elucidating both the spin and lattice dynamics of this metallic ferromagnet. We report a large measured value of the spin wave stiffness constant, which directly shows that the electron transfer energy of the *d* band is large. The spin dynamics, including magnetic critical scattering, demonstrate that this material behaves similar to other typical metallic ferromagnets such as Fe or Ni. The crystal structure is rhombohedral, as previously reported, for all temperatures studied (below ~425K). We have observed new superlattice peaks which show that the primary rhombohedral lattice distortion arises from oxygen octahedra rotations resulting in an $R\bar{3}c$ structure. The superlattice reflection intensities – which are very sensitive to structural changes – are independent of temperature demonstrating that there is no primary lattice distortion anomaly at the magnetic transition temperature, $T_C$ = 378.1 K, however there is a lattice contraction.

## INTRODUCTION

The correlation between magnetism and conductivity in $La_{1-x}Sr_xMnO_3$ is a classic subject extensively studied during the 1950's and 60's [1]. Undoped (x=0), $LaMnO_3$ is an antiferromagnetic insulator. Upon doping with Sr, this perovskite oxide becomes a ferromagnetic metal; the change of magnetism was well explained by the double exchange hopping mechanism [2]. Since the discovery of the high temperature superconducting copper oxides, study of these hole-doped manganese oxides has seen a revival. This renewed interest is due partly to the fact that $LaMnO_3$ is a charge gap insulator [3] associated with the large correlation energy of the *d* electrons in the $e_g$ band, and also to the discovery of giant magnetoresistance phenomena in samples with Sr dopant densities in the $0.2 \le x \le 0.4$ regime [4,5]. The structural phase transition can also be induced by magnetic fields when the doping level is near $x \approx 0.17$ [6].

In $LaMnO_3$, the four *3d* electrons on the Mn site ($Mn^{3+}$) share both the lower $t_{2g}$ band and the higher $e_g$ band ($3d^4$:$t_{2g}^3 e_g^1$). Due to a strong Hund coupling, all the spins are parallel on a given Mn site. The electrons in the $e_g$ band are highly correlated, creating the large correlation gap above the Fermi energy. Sr dopants naturally induce holes in the $e_g$ band near the Fermi energy, eventually producing mobile holes and conduction. At some hole doping level, the ordering changes from antiferromagnetic to ferromagnetic with a large effective magnetic moment since the intra-atomic Hund coupling is strong enough [7] to align the spins within the same ionic site. Due to the strong intra-atomic ferromagnetic coupling between the itinerant holes and the localized spins (S=3/2), the system becomes a ferromagnet upon doping with holes. The current understanding of the Sr-doped systems is similar to the double exchange mechanism [2], but also includes a highly correlated electron system's metal-insulator transition [8] or a strong electron-phonon interaction [9], which relate back to the high-$T_C$ copper oxide materials.

Tokura and co-workers recently observed a giant negative change in the magnetoresistance and a substantial change in the magnetostriction near the Curie temperature ($T_C$) for an x ≈ 0.175 crystal [5]. Extensive experimental studies have since been undertaken exploring a wide range of Sr doping in this manganese oxide system [10,11]. One expects that not only the electronic parameters such as the band filling, the strength of the electron correlations and so forth, but also other dynamical properties in both spin and lattice motions (or their interplay), play a crucial role in the manifestation of the interesting bulk properties. In other words, the $La_{1-x}Sr_xMnO_3$ system provides an opportunity to study a number of important physical problems: the metal-insulator electronic transition, the antiferromagnetic-ferromagnetic ordering phase change, and the lattice dynamics in the perovskite structure associated with these phase transitions.

The spin dynamics of $La_{0.7}Sr_{0.3}MnO_3$ crystals, information crucial for determining the itineracy of the system, have not been adequately studied. In $La_{1-x}Sr_xMnO_3$ (for x≈0.3) the inter-atomic exchange integrals, or the ferromagnetic coupling constant between nearest neighbor manganese atom spins, is predicted to be very strong due to the large transfer energy from holes hopping in the $e_g$ band [12]. This exchange constant can be directly extracted from neutron scattering measurements of the spin-wave stiffness constant.

In the present publication, we describe measurements of the dynamical magnetic properties and show that the spin wave stiffness is large, indicating a large transferred energy and itinerant ferromagnetic behavior. In addition, we describe ongoing structural measurements which confirm that the crystal structure is rhombohedral, as previously reported [1]; we observe a set of superlattice reflections from which we determine this primary structural distortion is due to shifts of the oxygen atoms and it has no structural anomaly at the Curie temperature.

## EXPERIMENT

The single crystal used in the present neutron scattering measurement was grown at the Joint Research Center for Atom Technology using the floating zone (FZ) method, as described in detail previously [10]. The crystal size is 3-4 mm in diameter and 8 cm in length. The largest domain of the crystal, which occupies about 1/3 of its length, was lined up such that the $[1\bar{1}1]$ and [011] axes lie in the scattering plane (the $[21\bar{1}]$ axis is vertical). We used the H8 and H9 triple axis spectrometers at the High Flux Beam Reactor at Brookhaven National Laboratory; H8 is a thermal (typically E = 14.7 or 30.5meV) neutron source, H9 provides a cold (E = 5meV) neutron beam. The (002) reflection of pyrolytic graphite crystals were used as monochromator and analyzer. Collimations were 40'-40'-S-80'-80' for H8 and 60'-40'-30'-S-80'-80' for H9 for studying the magnetism. Structural investigations were carried out at these collimations, as well as using a higher resolution collimation of up to 10'-10'-S-10'-80' at H8.

Figure 1 shows the phase diagram for $La_{1-x}Sr_xMnO_3$. In the $0.15 \le x \le 0.3$ dopant regime, a phase transition from ferromagnetic insulator to ferromagnetic metal as well as from orthorhombic to rhombohedral crystal structure occurs [6,11]. We decided to begin our investigations at higher dopant level ($x \approx 0.3$), in part due to the availability of large single crystals, and also in an effort to first understand the dynamics of a less complicated system before tackling the many interesting transitions occurring in the $0.15 \le x \le 0.3$ samples. At $x \approx 0.3$ the sample is a ferromagnetic metal at room temperature with a Curie temperature near 380K. Above $T_C$ the sample is a metallic paramagnet. The inset to Figure 1 shows the resistivity measured on this crystal indicating that the transition from paramagnet to ferromagnet is accompanied by a large drop in the resistivity, while maintaining an overall metallic temperature dependence.

MAGNETISM

Figure 2 shows the intensity of the (011) Bragg reflection as a function of temperature. This reflection has a particularly weak nuclear structure factor, $<F_N>^2 = 0.035 \times 10^{-24}$ cm$^2$, and therefore has a small intensity in the paramagnetic phase. Below $T_C$, magnetic scattering due to the ferromagnetism of spins aligning on the manganese atoms will produce a magnetic structure factor, $<F_M>^2 = 2/3\ (e^2\gamma / mc^2)^2 S^2 f^2$, where the 2/3 is because we are averaging over magnetic domains with different spin directions, the term in parentheses is a constant, $(e^2\gamma / mc^2)^2 = 0.538$, the form factor f is known for Mn at this wavevector to be ~0.6, and $S = \mu / 2$ which is to be determined [14]. The total scattering far below $T_C$ is given by $<F_N>^2 + <F_M>^2$. Using the measured peak intensities above and far below the Curie temperature from Fig. 2, we solve for a magnetic moment per Mn atom of $\mu \approx 3.6\mu_B$. This value is close to that obtained for the similar material $La_{1-x}Ca_xMnO_3$ at $x \approx 0.3$, $\mu \approx 3.5\mu_B$, by Wollan and Koehler [1]. It is also interesting to note that this crystal of $La_{0.7}Sr_{0.3}MnO_3$ is a rare example of a ferromagnetic oxide with a large magnetic moment ($\mu_{Mn} \leq 4\ \mu_B$) which shows metallic properties.

The inset of Fig. 2 shows the data near $T_C$, and a fit to a power law. The best fit gives $T_C = 378.1$ K and an exponent of $2\beta = 0.59 \pm 0.004$. The resultant exponent, $\beta = 0.295$, is near the well known 3-d Heisenberg ferromagnet model value of 1/3.

We also measured the characteristics of the spin wave in the ferromagnetic state to determine how this material compares to other ferromagnets. Figure 3 plots the dispersion curve of the spin wave away from the (011) point for two temperatures ($T = 300$ K $= 0.79\ T_C$ and $T = 27$ K $= 0.07\ T_C$). The dispersions are well fit by parabolic functions in q,

$$E = E_0 + Dq^2 , \tag{1}$$

where $E_0$ is the spin wave energy gap and D is the spin wave stiffness constant. The fitted values of D (shown in Fig. 3) are 188 ± 8 meVÅ$^2$ and 114 ± 4 meVÅ$^2$ for 27 K and 300 K, respectively. This change in D with temperature is very similar to previous measurements made on Fe at very low temperatures (D = 320 meVÅ$^2$) and at 0.8 $T_C$ (D = 175 meVÅ$^2$) [13]. Although energy scans at constant q = (011) did not reveal a spin wave energy gap at either temperature within our resolution limit, the best fits to the data in Figure 3 required small values of $E_0$: 0.75 ± 0.40 meV and 0.35 ± 0.22 meV for T = 27 K and 300 K, respectively. Higher resolution experiments are needed to determine if the spin-wave energy gap is indeed non-zero.

Another important topic for study in any ferromagnetic system is the paramagnetic scattering ($T > T_C$) [13]. Constant-q scans should reveal a Lorentzian function centered at zero energy transfer with a characteristic energy width having a functional form of

$$\Gamma(q) = A\, f\left(\frac{\kappa_1}{q}\right) q^{2.5}\ , \tag{2}$$

where A is a constant [13], $\kappa_1$ is the inverse correlation length, and $f(\kappa_1 / q)$ is the Résibois-Piette function [15] (which is unity at $T_C$ and becomes proportional to $(\kappa_1 / q)^{1/2}$ in the hydrodynamic region). $\kappa_1$ is given by $\kappa_0\, t^{0.7}$ where t the reduced temperature, ($t = T/T_C$). A is related to the spin-stiffness constant D [16]; when using the same system of units (meV and Å to the appropriate power) A and D have experimentally been found to be of the same order of magnitude. In typical metallic ferromagnets such as iron and nickel, D and A are both large [13].

We plot a constant-q scan for $La_{0.7}Sr_{0.3}MnO_3$ at T = 408K = 1.08$T_C$ which shows a typical Lorentzian line shape in Figure 4(a). As has been shown previously [13], Figure 4(b) displays a constant-E scan illustrating a profile that resembles spin-waves. From these paramagnetic studies, we have made preliminary measurements of A and obtain a value of A ~ 100meVÅ$^{2.5}$. This value is large and comparable in size to D.

STRUCTURE

Earlier measurements of $La_{0.7}Sr_{0.3}MnO_3$ [1] have shown that its structure is nearly cubic, with a rhombohedral distortion. This distortion can be observed by the splitting of the (hhh) peaks along the longitudinal direction (due to different domains), as well as from an accurate alignment of the [h00] and [0kk] axes. In Figure 5(a) we plot the high-resolution longitudinal scan through the (111) Bragg peak at room temperature. From the observed splitting we determine a value of $\alpha = 90.46 \pm 0.03^\circ$; we obtain the same $\alpha$ from measuring the angle between [100] and [011]. We measure the lattice constant to be a = 3.876 ± 0.003Å at room temperature.

A preliminary structural study has revealed superlattice peaks at 1/2(hkl), (with h, k, and l all odd - one such peak is plotted in Fig. 5(b) ), and missing superlattice peaks observed for 1/2(hhh) (with h odd). These superlattice peaks were confirmed in a powder sample study, one such $2\theta$ scan is shown in Figure 6. The intensities of these half-integer points, and the missing intensities of the 1/2(hhh) superlattice points, are well fit by a simple model for the structure factor of such points given by Axe *et al.* [17] for $LaAlO_3$. We can therefore see that the distortion is due to rotations of the $MO_6$ oxygen octahedra along each (111) axis to obtain a rhombohedral symmetry or an $R_{25}$ phonon condensation. The point group is $R\bar{3}c$, one of the most common distortions in perovskite systems [18], with a rhombohedral unit cell containing two $ABC_3$ units [19]. Rietveld refinement [20] was carried out on the powder spectrum of Figure 6 with the structure shown in the inset of Fig. 6. The oxygen atoms shift in the directions of the arrows by 0.24 ± 0.01Å. The Mn - O bond lengths are 1.960Å and the O-Mn-O angle is either 90.7˚ or 89.3˚.

The inset to Fig. 5(b) demonstrates that the distortion indicated by these superlattice reflections is unchanged when passing through the Curie temperature (compare this to the large change due to the onset of magnetic scattering observed for the

(011) Bragg peak in Figure 2). Rietveld fits of 300 K and 400 K powder reflection data also show the oxygen shifts change by less than our measurement error bar of 0.01Å. Therefore this primary rhombohedral structural distortion is *not* related to the onset of ferromagnetism. If oxygen octahedra rotations are the only distortion in this material, we can rule out Jahn-Teller distortions since they freeze the rotation of the octahedra along a specific crystallographic axis at the phase transition temperature. A secondary lattice distortion, a lattice contraction of 0.1% when cooling from T = 400 K to 300 K, is observed upon cooling and may be related to a Jahn-Teller effect. More detailed synchrotron structural studies are in progress to determine if further lattice distortions are present in the $La_{0.7}Sr_{0.3}MnO_3$ system.

## DISCUSSION

For localized ferromagnetic systems, the energy of the spin wave near the zone boundary is approximately equal to $k_BT_C$. However in $La_{0.7}Sr_{0.3}MnO_3$, we find that the spin wave energy is significantly larger than the Curie temperature of 378.1 K, as demonstrated by the large value of D. Our neutron scattering measurements determining a large D (and A) therefore indicate that this system not localized, but instead is itinerant in character. The renormalization of $T_C$ from the mean-field value by a Heisenberg Hamiltonian (appropriate for ferromagnetic insulators or localized spin systems) is considered to be a good measure of the itineracy; an itinerant ferromagnet will have a lower $T_C$ compared to the mean-field value, but will have large values of A and D [21], consistent with the present results. The strong transfer energy is consistent with a moderate $T_C$ in this itinerant system, and reflects a large J (favoring a t-J model). We have also concluded from comparing D with $T_C$ that, even with a strong measured magnetic moment ($3.6\mu_B$), the ferromagnetism in $La_{0.7}Sr_{0.3}MnO_3$ is itinerant in character.

A recent paper by Millis *et al.* [9] argues that a polaron effect from very strong electron-phonon coupling leads to better agreement with transport results in $La_{1-x}Sr_xMnO_3$ for $0.2 \le x \le 0.4$. Millis *et al.* also point out that such a polaron picture requires that the fluxuating oxygen displacements should be large (of the other of 0.1Å) for temperatures above $T_C$, and small for $T < T_C$. Our structural results demonstrate that the primary lattice distortion arising from $R_{25}$-type oxygen displacements is independent of the temperature being above or below $T_C$. In particular, the average oxygen atoms positions are the same within 0.01 Å at T = 400 K and 300 K. It remains to be seen in secondary lattice distortions, such as the observed lattice contraction, are related to polarons or if we can rule out a polaron picture at least for the x=0.3 compound. This still leaves the unresolved problem that double-exchange mechanisms alone do not entirely describe the properties of these novel materials [9].

## ACKNOWLEDGMENTS

We would like to thank R.J. Birgeneau, D.E. Cox and P. Böni for enlightening discussions. This work was supported in part by the U.S.- Japan Cooperative Program on Neutron Scattering and by the Ministry of Education, Science and Culture, Japan. Research at Brookhaven National Laboratory was supported by the Division of Materials Research at the U.S. Department of Energy, under contract No. DE-AC02-76CH00016. Work at Tohoku University was supported by the Grant-in-Aid for the Priority Area. Research at the Joint Research Center for Atom Technology was supported by the New Energy and Industrial Technology Development Organization (NEDO) of Japan.

## FIGURES

Figure 1. Electronic and magnetic phase diagram of $La_{1-x}Sr_xMnO_3$. Inset shows resistivity of the x=0.3 crystal used in the present study. (From Urushibara *et. al* [11]).

Figure 2. Intensity of the (011) Bragg peak as a function of temperature. Inset shows the data (open squares) near the Curie temperature and a fit (solid line) to a power law. The estimated magnetic moment of the Mn ions in the ferromagnetic state is $3.6\mu_B$.

Figure 3. Spin-wave dispersion along (011) at 300K (solid circles) and 27K (open circles). Horizontal lines indicate constant-energy scans; vertical lines indicate constant-q scans. Solid lines are fits to $E = E_0 + Dq^2$, where $E_0$ is the spin-wave energy gap and D is the spin-wave stiffness constant. The values of $E_0$ come from fits to the higher q data, however no spin-wave energy gaps were directly observed at q=0.

Figure 4. A constant-q scan (a) and a constant-E scan (b) for $La_{0.7}Sr_{0.3}MnO_3$ in its paramagnetic state ($T > T_C$).

Figure 5. (a) Splitting of the (111) peak and (b) a superlattice peak at $\frac{1}{2}(133)$ due to the rhombohedral distortion, both measured at room temperature. Inset to (b) illustrates that this structural distortion does not change when passing through $T_C$.

Figure 6. Powder reflection scattering data obtained using 30.5 meV neutrons; the solid line connecting the data points is a guide to the eye and peaks are labels with the cubic indices for each reflection. The $R\bar{3}c$ structure is drawn in the inset, with arrows indicating the displacements of the oxygen atoms away from their face-centered positions.

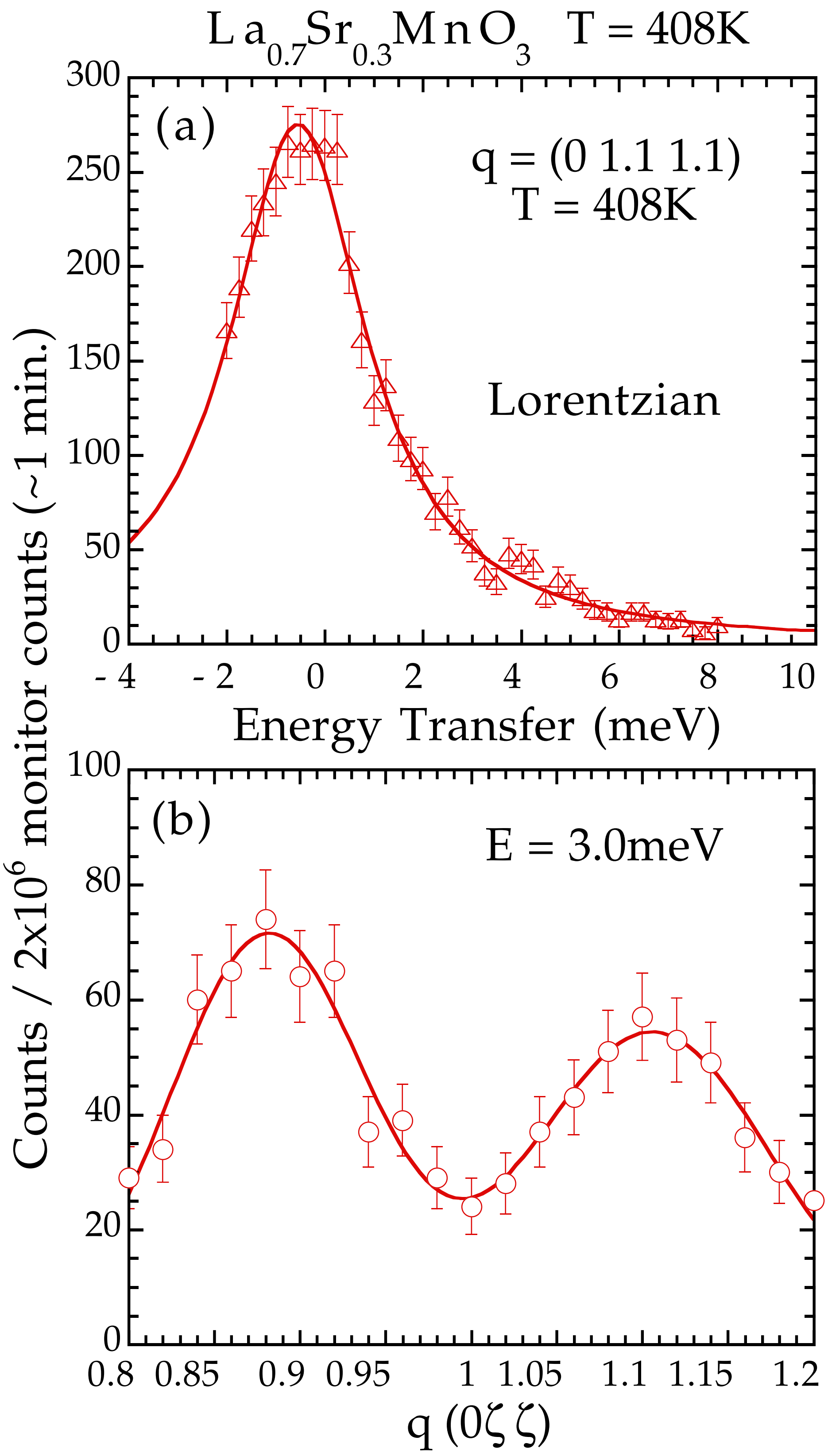

M.C. Martin et al., Figure 4

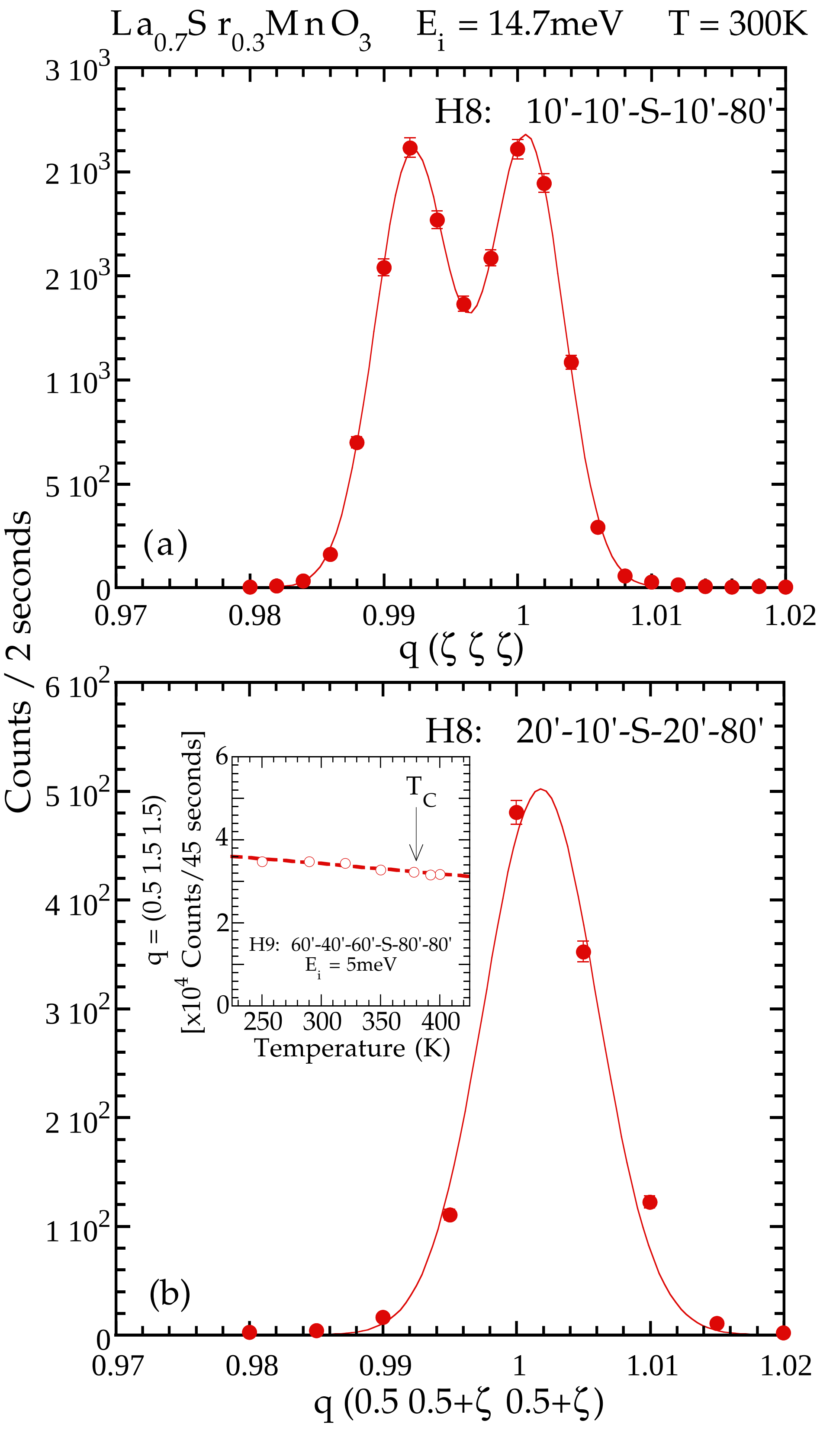

M.C. Martin et al., Figure 5

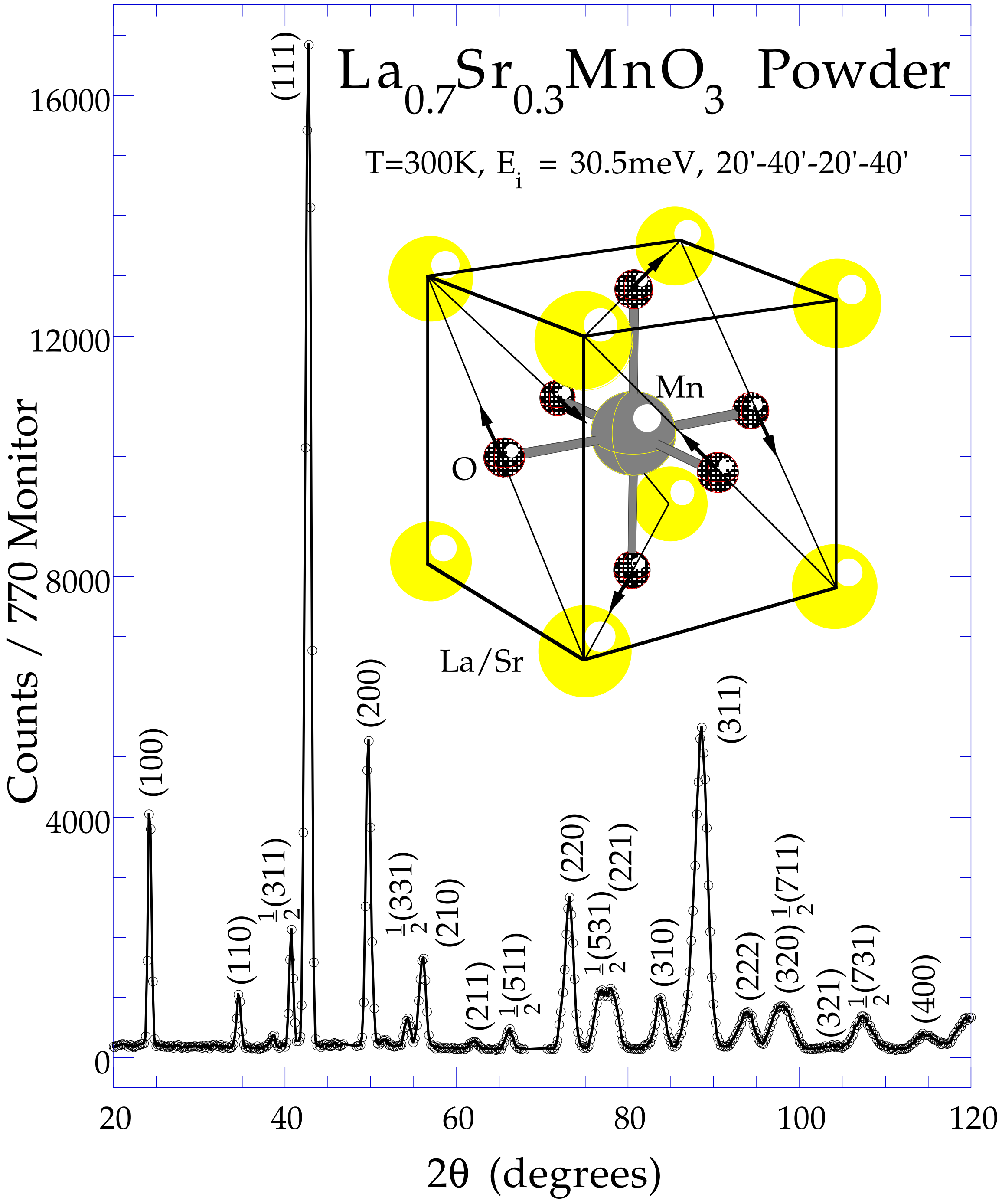

Martin et al., Figure 6